\documentclass[conference,9pt]{IEEEtran}
\IEEEoverridecommandlockouts

\usepackage{cite}
\usepackage{amsmath,amssymb,amsfonts,bm}
\usepackage{algorithmic}
\usepackage{graphicx}
\usepackage{textcomp}
\usepackage{xcolor}
\usepackage{multirow}
\usepackage{booktabs,url}
\usepackage{cleveref}

\newcommand{\setR}{\mathbb{R}}

\newcommand{\T}{\mathsf{T}}

\NewDocumentCommand\newletter{m m o m m}{
\NewDocumentCommand#1{s t@ o}{%
\IfBooleanTF{##1}{\mathbf{\MakeUppercase{#2}}\IfValueT{#3}{^{#3}}}{%
\IfBooleanTF{##2}{\mathbf{#2}\IfValueT{#3}{^{#3}}_{\IfValueTF{##3}{##3}{#5}}}{%
{#2}\IfValueT{#3}{^{#3}}_{\IfValueTF{##3}{##3}{#4}}%
}}}}

\NewDocumentCommand\newletterpre{m m m o m m}{
\NewDocumentCommand#1{s t@ o}{%
\IfBooleanTF{##1}{#2{\mathbf{\MakeUppercase{#3}}}\IfValueT{#4}{^{#4}}}{%
\IfBooleanTF{##2}{#2{\mathbf{#3}}\IfValueT{#4}{^{#4}}_{\IfValueTF{##3}{##3}{#6}}}{%
#2{{#3}}\IfValueT{#4}{^{#4}}_{\IfValueTF{##3}{##3}{#5}}%
}}}}

\NewDocumentCommand\newletterbm{m m o m m}{
\NewDocumentCommand#1{s t@ o}{%
\IfBooleanTF{##1}{\bm{\MakeUppercase{#2}}\IfValueT{#3}{^{#3}}}{%
\IfBooleanTF{##2}{\bm{#2}\IfValueT{#3}{^{#3}}_{\IfValueTF{##3}{##3}{#5}}}{%
{#2}\IfValueT{#3}{^{#3}}_{\IfValueTF{##3}{##3}{#4}}%
}}}}

\def\R{\mathbb{R}}

\newletter{\x}{x}{t,d}{t}
\newletter{\px}{x^\prime}{t,d}{t}
\newletter{\stdx}{x}[\text{(std)}]{t,d}{t}

\newletter{\eigv}{V}{d,d'}{d}
\newletter{\eigw}{\Lambda}{d,d'}{d}
\newletter{\xpca}{x}[\text{(pca)}]{t,d}{t}
\newletter{\xwht}{x}[\text{(whiten)}]{t,d}{t}
\newletter{\cov}{c}{d,d'}{d}

\newletter{\s}{s}{t,d}{t}
\newletter{\amix}{a}{d,d'}{d}
\newletter{\w}{w}{d,d'}{d}
\newletter{\y}{y}{t,d}{t}
\newletter{\xica}{x}[\text{(ica)}]{t,d}{t}

\begin{document}

\title{Discrete Speech Unit Extraction\\ via Independent Component Analysis
\thanks{
This study was supported by the BRIDGE program of the Cabinet Office, Government of Japan.
This work used the Bridges2 system at PSC and Delta system at NCSA through allocation CIS210014 from the Advanced Cyberinfrastructure Coordination Ecosystem: Services \& Support (ACCESS) program, which is supported by National Science Foundation grants \#2138259, \#2138286, \#2138307, \#2137603, and \#2138296.}
}


\author{
    \IEEEauthorblockN{
        Tomohiko Nakamura\thanks{\IEEEauthorrefmark{1}Equal contribution.}\IEEEauthorrefmark{1}\IEEEauthorrefmark{4}, %
        Kwanghee Choi\IEEEauthorrefmark{1}\IEEEauthorrefmark{2}, %
        Keigo Hojo\IEEEauthorrefmark{4}\IEEEauthorrefmark{3}, %
        Yoshiaki Bando\IEEEauthorrefmark{4}, %
        Satoru Fukayama\IEEEauthorrefmark{4}, %
        and Shinji Watanabe\IEEEauthorrefmark{2}\vspace{.0\baselineskip}}
    \IEEEauthorblockA{
        \IEEEauthorrefmark{4}National Institute of Advanced Industrial Science and Technology, Japan}
    \IEEEauthorblockA{
        \IEEEauthorrefmark{2}Carnegie Mellon University, USA, 
        \IEEEauthorrefmark{3}Toyohashi University of Technology, Japan}
}

\maketitle

\begin{abstract}
Self-supervised speech models (S3Ms) have become a common tool for the speech processing community, leveraging representations for downstream tasks.
Clustering S3M representations yields discrete speech units (DSUs), which serve as compact representations for speech signals.
DSUs are typically obtained by $k$-means clustering.
Using DSUs often leads to strong performance in various tasks, including automatic speech recognition (ASR).
However, even with the high dimensionality and redundancy of S3M representations, preprocessing S3M representations for better clustering remains unexplored, even though it can affect the quality of DSUs.
In this paper, we investigate the potential of linear preprocessing methods for extracting DSUs. 
We evaluate standardization, principal component analysis, whitening, and independent component analysis (ICA) on DSU-based ASR benchmarks and demonstrate their effectiveness as preprocessing for $k$-means.
We also conduct extensive analyses of their behavior, such as orthogonality or interpretability of individual components of ICA.
\end{abstract}


\section{Introduction}
\label{sec:intro}
Self-supervised speech models (S3Ms) \cite{baevski20wav2vec2,hsu21hubert,chen21wavlm,babu22xlsr} has become the prevailing approach for addressing a variety of speech and audio processing tasks.
These models have demonstrated significant effectiveness across numerous tasks, as evidenced by extensive benchmark evaluations \cite{yang21superb,tsai22superbsg,shi21mlsuperb,huang24dynamic}.
For evaluations, S3Ms are often frozen, and a small trainable layer (\textit{e.g.}, a single linear layer) is added and fine-tuned for a specific downstream task.  
This process, which is further corroborated by various linear probing literatures \cite{PasadCL21,PasadSL23,choi24mi}, highlights the efficacy of using frozen pre-trained S3Ms.

The self-supervised learning paradigm makes those frozen S3M representations ``useful'' in downstream tasks \cite{choi24mi}.
S3Ms are trained to reconstruct the hidden signals based on the neighboring context \cite{baevski20wav2vec2,hsu21hubert}.
Its loss depends on the representation-wise distances, nudging the representations of similar signals to be in proximity \cite{choi2022opening,choi24b_interspeech}.
Consequently, clustering similar signals based on their representation distances, or discrete speech units (DSUs), becomes a natural progression \cite{chang2024dsuchallenge,chang2024exploring}.

\begin{figure}
    \centering
    \includegraphics[width=\columnwidth]{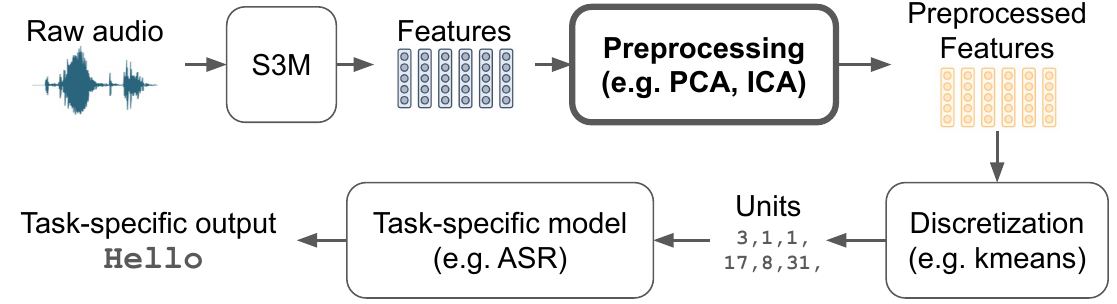}
    \caption{Schematic illustration of processing flow for DSU-based ASR model. Our work focuses on improving the preprocessing step for $k$-means (bold).}
    \label{fig:dasr_flow}
\end{figure}

In \Cref{fig:dasr_flow}, we show the typical use of DSUs for downstream tasks.
DSUs are typically derived by applying the $k$-means algorithm to S3M representations \cite{chang2024dsuchallenge}.
The resulting cluster indices, \textit{i.e.}, DSUs, can be viewed as the tokenized or summarized speech.
These units requires substantially less storage compared to conventional continuous features, such as spectral speech features \cite{chang2024exploring,borsos2023audiolm}.
They have demonstrated strong empirical performance across various tasks, including automatic speech recognition (ASR) \cite{chang2024exploring}, speech-to-speech translation (S2ST) \cite{LeeCWGPMPAHTPH22,InagumaPKCWC00023}, and speech synthesis \cite{borsos2023audiolm,chang2024dsuchallenge}.
Moreover, they facilitate unified modeling of text and speech more easily \cite{rubenstein2023audiopalm,maiti2024voxtlm}, or further, textless natural language processing (NLP) \cite{lakhotia2021generative,hassid2024textually}.

Accordingly, there has been growing interest in how to extract DSUs from S3Ms.
Many have shown that there is an optimal transformer layer \cite{chang2024exploring} and number of clusters \cite{van2022comparison}, both of which vary based on specific downstream tasks.
Also, normalizing each dimension of the representation to have zero mean and unit variance prior to $k$-means has been shown effective \cite{borsos2023audiolm}.
Finally, compressing the resulting DSUs through deduplication and byte-pair encoding (BPE) has been demonstrated to enhance computational efficiency while having minimal impact on downstream performance \cite{chang2024exploring}.

However, limited research has focused on enhancing $k$-means clustering with preprocessing techniques for extracting DSUs.
Hence, we investigate the potential of linear preprocessing methods for $k$-means \cite{liang2014improved}; an area yet to be explored in detail, despite the established effectiveness of these methods in other contexts.
For instance, principal component analysis (PCA) has been applied before $k$-means to reduce the computation cost \cite{liang2014improved}.
Also, PCA is used to extract spoken word embeddings from S3Ms \cite{sanabria2023analyzing} and analyze S3M representations \cite{tom2022wav2vec,mohamed2024orthogonality}, while independent component analysis (ICA) has been used for analyzing textual word embeddings \cite{YamagiwaOS23}.
We evaluate these linear transformations as preprocessing methods for clustering S3M representations to extract DSUs, utilizing a recently proposed benchmark \cite{chang2024dsuchallenge}.
Additionally, we conduct a qualitative analysis of their behaviors, focusing on the orthogonality and interpretability of individual components.

\section{Linear Preprocessing Methods \\ for Discrete Speech Unit Extraction} \label{sec:prelim}
\subsection{$k$-means and Linear Transformation}
Let $\x* \in \setR^{T\times D}$ be the concatenation of S3M features of the training data prepared for $k$-means, where $T$ and $D$ are the number of frames and the feature dimension.
The conventional DSU extraction method directly applies $k$-means to $\x*$:
\begin{equation}
    \{u_{t}\}_{t=1}^T=\text{kmeans}(\x*),
\end{equation}
where $u_{t}\in\{0,1,\ldots,k-1\}$ is the DSU of frame $t$.
It is important to note that the distance metric of $k$-means for DSU extraction (e.g., the Euclidean distance) is not invariant to arbitrary linear transformations on $\x*$.
In other words, linearly transforming $\x*$ can yield different DSUs.
This motivates an investigation into how linear transformations may affect the $k$-means-based DSU extraction.

\subsection{Linear Preprocessing Methods} \label{sec:preproc_methods}
We explore four methods: standardization, PCA, whitening, and ICA.
Each linear processing transforms input data based on statistics, and is trainable in an unsupervised manner.  

\textbf{Standardization:} Standardization normalizes each column of $\x*$ to have zero mean and unit variance.
Let $\x\in\R$ denote the $(t,d)$th entry of $\x*$, where $d$ is the feature dimension index.
The standardized feature $\stdx$ is given as:
\begin{equation}
    \stdx = \dfrac{1}{\sigma_d}(\x-\bar{x}_d),
\end{equation}
where $\bar{x}_d\in\R$ and $\sigma_d\in\R$ are the average and standard deviation of feature dimension $d$, respectively.
Standardization is used for preprocessing w2v-BERT\cite{chung2021w2vbert} features in \cite{borsos2023audiolm}.

\textbf{PCA:} PCA extracts correlated components from input data using the first- and second-order statistics.
It subtracts the average over samples from $\x*$ and obtain the mean-centered data $\px*$.
Then, eigenvalue decomposition of the covariance matrix $\cov*:=\px*^\T\px*/(T-1)$ is computed.
Since $\cov*$ is symmetric, it can be diagonalized as:
\begin{equation}
    \cov* = \eigv* \eigw*\eigv*^\T,
\end{equation}
where $\eigv* \in \setR^{D\times D}$ is the orthogonal matrix and $\eigw* \in \setR^{D\times D}$ is the diagonal matrix containing the eigenvalues on the main diagonal.
The principal components of the data is given as:
\begin{equation}
    \xpca* =\px* \eigv*. \label{eq:pca}
\end{equation}

\textbf{Whitening:} Whitening (a.k.a. sphering) further normalizes the variances of the principal components:
\begin{equation}
    \xwht* = \xpca* \eigw*^{-\frac{1}{2}}.
    \label{eq:whiten}
\end{equation}
Previous NLP works used this technique to extract sentence embeddings from pretrained models \cite{su2021whitening,huang2021whiteningbert}.

\textbf{ICA:} ICA has underpinned multichannel audio source separation\cite{sawada2019apsipatrans}. 
Unlike PCA, it finds statistically independent components using higher-order statistics.
It assumes that $\xwht*$ consists of $D$ independent components and is observed through the unknown mixing system $\amix*\in\R^{D\times D}$:
\begin{equation}
    \xwht* = \s* \amix*^\T,
\end{equation}
where $\s*\in \setR^{T\times D}$ represents the weights of the independent components.
The goal is to find the inverse system, characterized by the demixing matrix $\w*\in \setR^{D\times D}$, so that the separated components $\xwht* \w*^\top$ approximately equals $\s*$.

This problem can be formulated as the maximum likelihood estimation with respect to $\w*$ \cite{cardoso1997spl}.
The log-likelihood function is:
\begin{equation}
    \log p(\xwht*; \w*) = \log p(\s*=\xwht* \w*^\top) + 
        \log |\det \w*|.
\end{equation}
The probability distribution of $\s*$ determines the characteristics of the independent components, with the standard Laplace distribution typically chosen for each entry of $\s*$:
\begin{equation}
    \log p(\s*) = -\sum_{t,d=1}^{T,D}|\s| - TD\log 2,
\end{equation}
where $\s$ is the $(t,d)$th entry of $\s*$.
With this formulation, $\w*$ can be estimated using an iterative algorithm that ensures the log-likelihood function does not decrease at each iteration.
Details of this algorithm can be found in \cite{Ono2010LVAICA}.

Using the $\w*$ estimate, $\hat{\w*}$, the ICA-transformed data is given as:
\begin{equation}
    \xica* = \xwht* \hat{\w*}^\T.
    \label{eq:ica}
\end{equation}

\section{Experimental Results} \label{sec:exp}
\subsection{Experimental Setup} \label{sec:exp_cond}
We evaluated the effects of PCA, whitening, ICA on downstream tasks via DSU challenge~\cite{chang2024dsuchallenge}, which is the recently proposed challenge to provide a unified benchmark for DSUs.
We followed the ASR track, evaluating whether DSUs can capture phonetic information well for both rich (English) and low-resource (multilingual) scenarios.
Our code\footnote{\url{https://github.com/TomohikoNakamura/ica_dsu_espnet}} was implemented based on the official code of the DSU challenge\footnote{\url{https://github.com/espnet/espnet/tree/master/egs2/interspeech2024_dsu_challenge/asr2}}.

\begin{table}[t]
\centering
{
    \tabcolsep=0.4ex
    \caption{CERs [\%] and bit-rates [bit/s] on the DSU challenge test set when using $k$-means clustering with and without preprocessing methods for XLS-R-300M}
    \centering
    \begin{tabular}{cc|c|ccccc}
        \toprule
        Preproc. & $k$-means & bit-rate & dev\_clean & dev\_other & test\_clean & test\_other & test\_1h \\ \midrule
        \multirow{2}{*}{-} & Euclid. & 394 & 3.2 & 9.1 & 3.2 & 9.2 & 22.9 \\
         & Cosine & 398 & 3.3 & 9.1 & 3.2 & 9.2 & 22.7 \\ \midrule
        \multirow{2}{*}{Std~\cite{borsos2023audiolm}} & Euclid. & 398 & 2.7 & 7.6 & 2.6 & 7.6 & 21.7 \\
         & Cosine & 411 & 2.6 & 7.6 & 2.6 & 7.5 & \textbf{20.9} \\ \midrule
        \multirow{2}{*}{PCA} & Euclid. & 394 & 3.2 & 9.0 & 3.1 & 9.2 & 22.9 \\
         & Cosine & 401 & 3.3 & 9.3 & 3.4 & 9.3 & 22.8 \\ \midrule
        \multirow{2}{*}{Whiten} & Euclid. & \textbf{284} & 2.9 & 7.8 & 2.8 & 7.6 & 27.1 \\
         & Cosine & 373 & 2.6 & 7.3 & \textbf{2.5} & 7.2 & 21.3 \\ \midrule
        \multirow{2}{*}{ICA} & Euclid. & 296 & 2.8 & 7.7 & 2.6 & 7.6 & 27.3 \\
         & Cosine & 372 & \textbf{2.5} & \textbf{7.1} & \textbf{2.5} & \textbf{7.1} & \textbf{20.9} \\
        \bottomrule
    \end{tabular}
    \label{tab:xls_r_300m}
}
\end{table}

\textbf{Dataset:}
Same with the DSU challenge, we use LibriSpeech 100-hour subset (LibriSpeech-100) \cite{panayotov2015icassp}, which comprises English read speech, and the ML-SUPERB 1-hour (ML-SUPERB-1h) benchmark, which includes around 220-hour speech of 143 languages.
The train and test set is the combination of these two corpora.
The test set consists of \emph{dev\_clean, dev\_other, test\_clean} and \emph{test\_other} sets of LibriSpeech-100; and \emph{test\_1h} set of ML-SUPERB-1h.

\textbf{DSU extraction method:}
For the S3M, we chose the 17th layer of \emph{XLS-R-300M}~\cite{babu22xlsr}, which was used in the top-3 systems on the ASR track of the DSU challenge~\cite{li2024dsu,chang2024dsuchallenge}.

Regarding preprocessing of DSU extraction, we compared the conventional (i.e., no-preprocessing) method with \emph{standardization (Std)}, \emph{PCA}, whitening (\emph{Whiten}), and \emph{ICA}, which are described in \Cref{sec:preproc_methods}.
The demixing matrix of ICA was initialized with an identity matrix and the number of iterations was 100 for the ICA algorithm.
As the distance metric of the $k$-means, the Euclidean and cosine distances were used, which we referred to as \emph{Euclid} and \emph{Cosine}, respectively.
All combinations of the preprocessing methods and the distance criteria were evaluated as DSU extraction methods.
For the $k$-means training, 5\% of the training set was randomly chosen\footnote{Baseline used 15\% of the training set, but we empirically found that reducing it to 5\% provided similar ASR performances with $k$-means.}.
The number of clusters was 2000.

\textbf{Discrete ASR model:}
For all DSU extraction methods, we used the baseline system from the DSU challenge, as proposed in \cite{chang23b_interspeech}. This system utilizes the joint CTC/attention-based encoder-decoder architecture based on the E-Branchformer \cite{kim2023slt}.
After the repetition of DSUs were removed, BPE is applied with a vocabulary size of 3,000.
For the target languages, the BPE is applied with a vocabulary size of 6,500.
The network was trained for 100 epochs, using the Adam optimizer with 30,000 warm-up steps for the learning rate.

\textbf{Evaluation metrics:}
The evaluation metrics are character error rates (CERs) and bit-rate, following \cite{chang2024dsuchallenge}.
Both metrics were computed for every test set.
The bit-rate measures the efficiency of the discrete representation.
Given the DSU sequence of length $N$, it is defined as 
\begin{equation}
    B = \dfrac{N\log_2V}{U},
\end{equation}
where $V$ is the vocabulary size, i.e., $V=3,000$ in our settings, and $U$ is the utterance duration in seconds.
The bit-rate values were averaged over all test utterances.

\subsection{Effects of different preprocessing techniques} \label{sec:exp_preproc}
\Cref{tab:xls_r_300m} shows CERs using DSUs extracted with and without the preprocessing methods.
Euclidean distance had similar CERs with and without PCA because it is invariant to both centralization and orthogonal projection.
While Cosine distance is not invariant to centralization, the Cosine result was unchanged by PCA.
We suspect that it is due to transformer representations being anisotropic (i.e., occupying a narrow cone in the geometrical space) \cite{GodeyCS24} so the centralization did not greatly change the geometry.
We further explore this aspect in detail in \Cref{subsec:ortho}.

Where PCA shows little to no improvement, Std, Whiten, and ICA substantially enhanced the ASR performances.
Using Cosine rather than Euclidean further improves performance.
Despite their performance improvements, Std resulted in higher bit-rates with Cosine compared to the no-preprocessed case.
By contrast, Whiten and ICA effectively reduced the bit-rates.
ICA with Cosine achieved the best average ASR performance with the lower bit-rate, demonstrating the importance of adequate preprocessing.


Whiten and ICA works better with Cosine than Euclid, particularly on the test\_1h set.
A key characteristic of these two preprocessing methods is that they normalize the norms of input features.
This result suggests that the norms of the whitened features are less relevant to ASR performance.

\subsection{Effect of the number of clusters}
\begin{table}[t]
\centering
{
    \tabcolsep=0.4ex
    \caption{CERs [\%] of Euclidean $k$-means and cosine-distance-based $k$-means with ICA for various s number of clusters} 
    \begin{tabular}{cc|c|ccccc} \toprule
    Method & \#clusters & bit-rate& dev\_clean & dev\_other & test\_clean & test\_other & test\_1h \\ \midrule
    \multirow{4}{*}{Euclid} &  100 & 149 & 27.1 & 33.0 & 26.0 & 33.5 & 52.3 \\
                            &  500 & 265 & 5.8 & 12.7 & 5.7 & 12.9 & 29.4 \\
                            & 1000 & 326 & 4.0 & 10.3 & 4.0 & 10.5 & 25.4 \\
                            & 2000 & 394 & 3.2 & 9.1 & 3.2 & 9.2 & 22.9 \\ \midrule
           &  100 & 138 & 32.1 & 28.3 & 29.8 & 27.7 & 54.5 \\ 
    ICA+   &  500 & 244 & 4.3 & 8.7 & 4.1 & 8.9 & 27.8 \\
    Cosine & 1000 & 301 & 3.0 & 7.7 & 2.9 & 7.6 & 23.6 \\
           & 2000 & 372 & 2.5 & 7.1 & 2.5 & 7.1 & 20.9 \\ \bottomrule 
    \end{tabular}
    \label{tab:varying_clusters}
}
\end{table}

We varied the number of clusters to examine the effectiveness of ICA on different settings.
We compared the conventional Euclid with ICA-preprocessed Cosine (ICA+Cosine), which achieved the best ASR performance in \Cref{sec:exp_preproc}.

\Cref{tab:varying_clusters} shows that both methods performed better with more clusters.
Note that deduplication and BPE was applied to DSUs as in \Cref{sec:exp_preproc}.
As in \cite{wu2024codec}, the results exhibit the trade-off between bit-rate and ASR performance.
ICA+Cosine outperforms Euclid by a consistent margin, except for 100 clusters, demonstrating the stability of ICA with a sufficient number of clusters.

\subsection{Effects of S3M characteristics}
We also examined which factor of S3M is essential for ICA to work effectively.
We tested S3Ms that differ in training methodology and data.
The first S3M, \emph{Wav2vec2.0-Large}~\cite{baevski20wav2vec2}, was trained by the same training methodology as XLS-R-300M, but its training data comprise English data only (the LibriSpeech corpus~\cite{panayotov2015icassp}).
Since it has the same network architecture as XLS-R-300M, we extracted DSUs from its 17th layer.
The second S3M, \emph{multilingual HuBERT (mHuBERT)}~\cite{lee2022naacl}, was trained with the multilingual data but with a different training methodology from XLS-R-300M.
For DSU extraction, we used 1000 clusters from mHuBERT's 11th layer, following its application in S2ST\cite{lee2022naacl}.
The third S3M, \emph{WavLM-Large}~\cite{chen21wavlm}, was trained on English data only, with a different training methodology from XLS-R-300M.
DSUs were extracted from its 21st layer, as in the DSU challenge baseline.
The number of clusters was set to 2000 except for mHuBERT, and other hyperparameters followed \Cref{sec:exp_cond}.

\Cref{tab:ablation} shows CERs obtained with Euclid and ICA+Cosine for the three S3Ms.
For Wav2vec2.0-Large, ICA+Cosine provided significant improvements for the LibriSpeech-100 test sets.
The cause of the performance degradation for the test\_1h set may be the mismatches between the training data of the S3M and downstream task.
Despite the improvements for Wav2vec2.0-Large, ICA+Cosine had higher CERs from Euclidean for the other S3Ms.
These results indicate that the effectiveness of ICA may depend on the underlying S3M training methodology, which is based on the masked language modeling and codebook groupings.
We aim to conduct further exploration on the impact of training methodology on S3M representation geometry in the future.

\begin{table}[t]
\centering
{
    \tabcolsep=0.4ex
    \caption{CERs [\%] for various S3Ms}
        \begin{tabular}{c|c|ccccc} \toprule
            S3M & Method & dev\_clean & dev\_other & test\_clean& test\_other & test\_1h \\ \midrule
            Wav2vec2.0-                     & Euclid& 2.3 & 6.1 & 2.4 & 6.1 & \textbf{27.2}\\
              Large \cite{baevski20wav2vec2}& ICA+Cosine & \textbf{1.7} & \textbf{4.0} & \textbf{1.7} & \textbf{3.9} & 30.4 \\ \midrule
            {mHuBERT}                       & Euclid & \textbf{5.6} & \textbf{12.5} & \textbf{5.5} & \textbf{12.5} & \textbf{31.2} \\
                                 \cite{lee2022naacl} & ICA+Cosine & 7.6 & 13.4 & 7.2 & 13.5 & 33.5 \\ \midrule
               WavLM-                & Euclid & \textbf{1.5} & \textbf{3.4} & \textbf{1.5} & \textbf{3.3} & \textbf{22.5} \\
            Large \cite{chen21wavlm} & ICA+Cosine & 1.8 & 3.8 & 1.7 & 3.8 & 24.8 \\
            \bottomrule
        \end{tabular}
    \label{tab:ablation}
}
\end{table}



\section{Qualitative Analysis of Clustering Results}
\subsection{Orthogonality of $k$-means centroids}\label{subsec:ortho}
\begin{figure}[t]
    \centering
    \includegraphics[width=0.9\columnwidth]{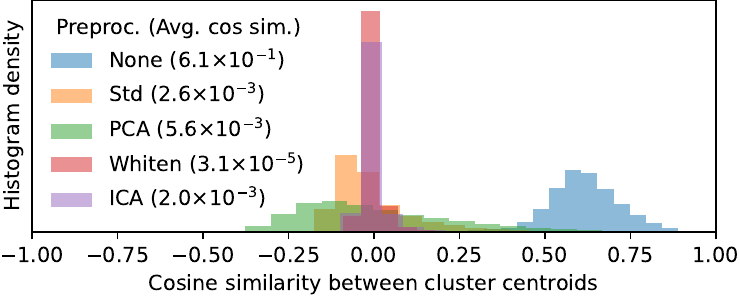}
    \caption{
    Comparing cosine similarities and their averages between $k$-means cluster centroids.
    Different preprocessing methods are used before clustering.
    }
    \label{fig:dists}
\end{figure}
To compare the cosine similarities between clustering results, we plot the histogram of the similarities between different preprocessing methods in \Cref{fig:dists}.
The $k$-means centroids were obtained with Cosine.
Without any preprocessing, clusters tend to be similar to each other, centering around 0.6.
As mentioned before, the anistropy of transformer representations \cite{GodeyCS24} also appears to hold for S3Ms.
Previous NLP works tried to make the representations more isotropic (opposite of anisotropic) to improve performance, such as word \cite{mu2018all,raunak2019effective} and sentence embeddings \cite{su2021whitening,huang2021whiteningbert}.
We can observe that all preprocessing methods significantly increase the isotropy of representations, yielding orthogonal cluster centroids.

\subsection{Interpretability of $k$-means centroids and ICA components}
To further understand the characteristics of $k$-means and ICA, we conducted a qualitative analysis of their behaviors, using the final layer representations of WavLM-Large.

\textbf{Dataset:} We used the TIMIT dataset \cite{garofolo1993darpa}, a phonetically transcribed English speech dataset.
We segmented the representation according to timestamps and applied average pooling to obtain phone-wise representations, similar to \cite{PasadCL21,PasadSL23}.
This yielded 187,737 phone utterances across 51 phones (excluding silence) from 630 speakers.

\begin{figure}[t]
    \centering
    \includegraphics[width=\columnwidth]{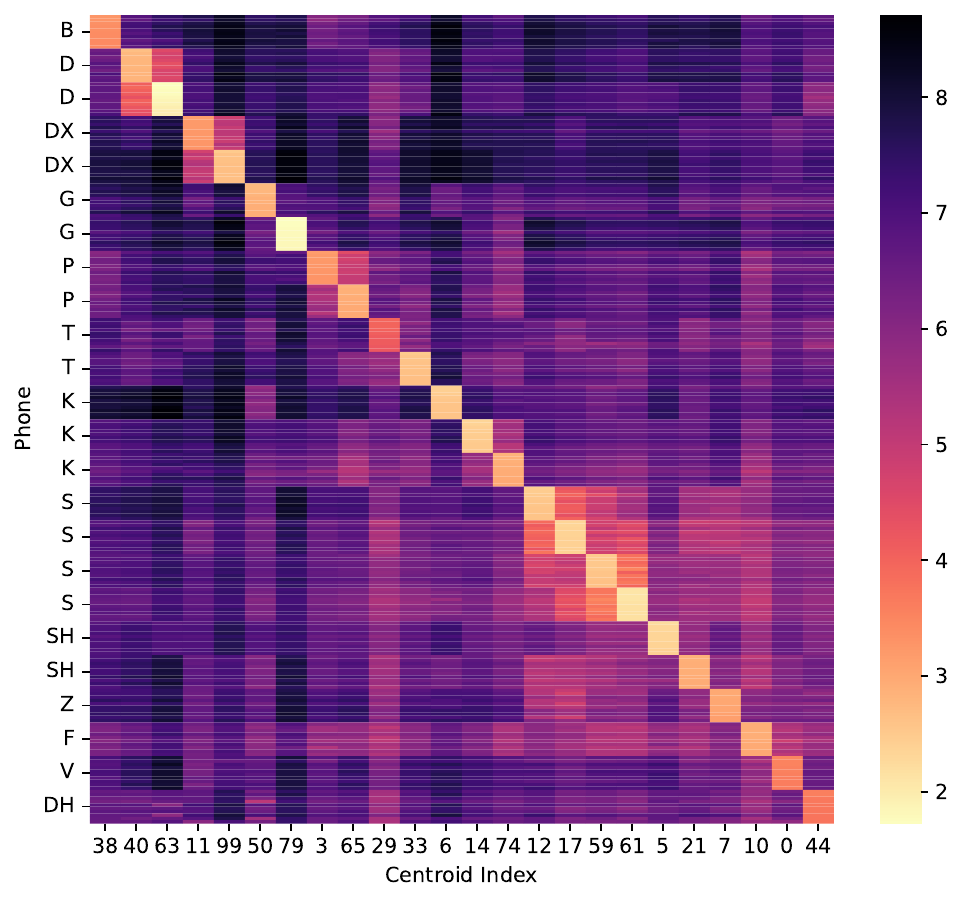}
    \caption{
    Comparing $k$-means centroids with phones with $k=100$.
    We plot the centroids where the 10 nearest neighbor representations are from the same phone.
    Due to space limitations, we plot stops and fricatives only.
    }
    \label{fig:kmeans-heatmap}
\end{figure}
\textbf{$k$-means centroids:} We examined the 10 nearest neighbors of each cluster centroid in \Cref{fig:kmeans-heatmap}, setting the cluster count to 100.
To simplify, we focus on cases where the top 10 nearest neighbor representations are from the same phone.
Due to space limitations, we plot only the stops and fricatives.

Consistent with previous work \cite{baevski20wav2vec2,hsu21hubert}, we find that clusters align well with phonetic labels and distinguish allophones (different phonetic realizations of same phoneme), such as [th] in ``thin'' and [dh] in ``than.''
Further, multiple clusters often map to a single phone, with similar phones like [D], [DX], and [S] clustering near each other, while others like [G] and [K] are more dispersed.
This suggests the clustering captures nuanced distinctions beyond traditional allophones.
Finally, fricatives ([S], [SH], [Z], [F], [V], [DH]) cluster closely, supporting prior findings \cite{abdullah23_interspeech,sicherman2023analysing} that representations of natural classes tend to be proximal.

\begin{figure}[t]
    \centering
    \includegraphics[width=0.9\columnwidth]{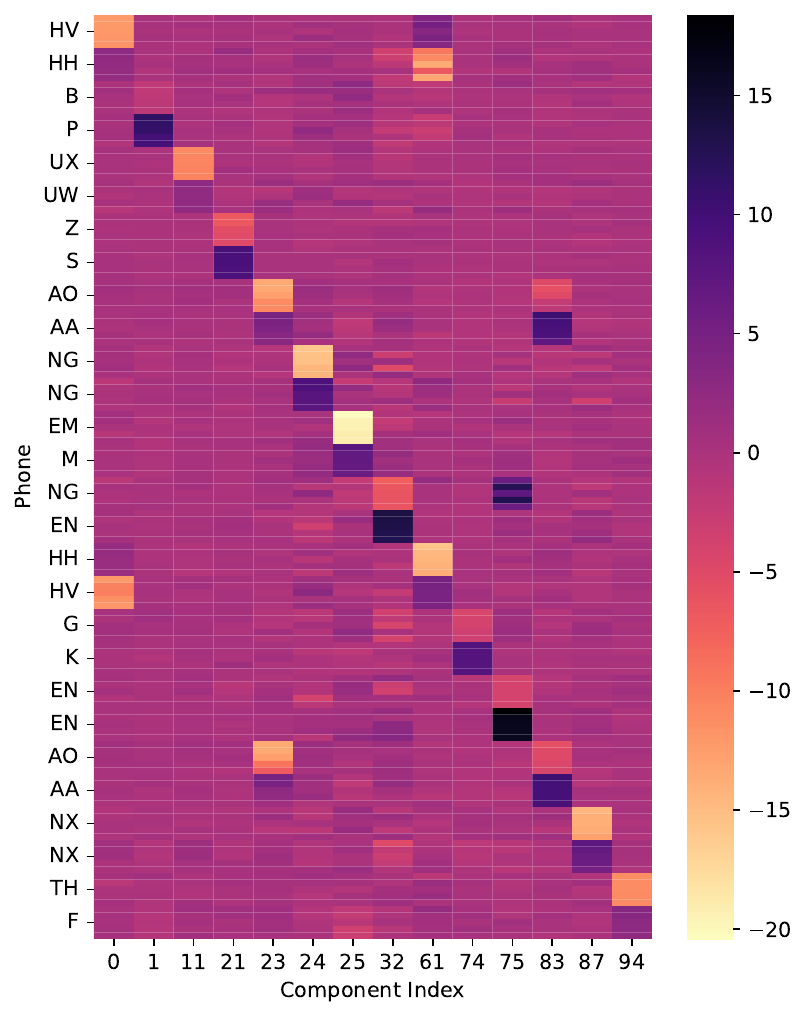}
    \caption{
    Comparing ICA components with phones where number of components is 100.
    We plot the components where the top 5 and bottom 5 representations are from the same phone.
    }
    \label{fig:ica-heatmap}
\end{figure}
\textbf{ICA components:} Unlike $k$-means centroids, ICA yields components with both positive and negative directions.
Hence, we examined both top 5 and bottom 5 representations for each component in \Cref{fig:ica-heatmap}, where the extremes are of the same phone.
We identify three interpretable component types.

The first type shows phonetic contrasts\footnote{In consonants, manner of articulation, place of articulation, and voicing are the three fundamental dimensions.}, such as voicing differences in component 1 ([B]/[P]), 21 ([Z]/[S]), and 61 ([G]/[K]), while preserving manner and place of articulation.
Similarly, component 94 ([F]/[TH]) contrasts nearby places of articulation.
The second type captures allophones of a single phoneme, seen in components 0 and 61 ([HH]/[HV]), 11 ([UX]/[UW]), 23 and 83 ([AO]/[AA]), and 25 ([EM]/[M]).
The third type includes repeated instances of the same phone within one component, similar to observations from $k$-means, indicating that S3M representations may capture fine acoustic distinctions beyond allophones.


In summary, we reconfirm that $k$-means effectively clusters phonetically similar sounds, where ICA goes a step further.
Not only does ICA uncovers similar sounds, but it also finds axes with linguistically interpretable contrasts, offering a deeper level of analysis.

\section{Conclusion}
We investigate the impact of the preprocessing methods for extracting DSUs from S3Ms.
Our experiments indicate that applying whitening and ICA before $k$-means clustering can improve the discrete ASR performances for XLS-R-300M with less bit-rates of DSUs.
Further, our analyses show the orthogonality and interpretability of ICA components.
Future work can be done for wider downstream tasks, such as vocoders, or further exploiting each ICA component.



\vfill\pagebreak

\bibliographystyle{ieeetr}
\bibliography{refs}

\begin{thebibliography}{10}

\bibitem{baevski20wav2vec2}
A.~Baevski, Y.~Zhou, A.~Mohamed, and M.~Auli, ``wav2vec 2.0: {A} framework for self-supervised learning of speech representations,'' in {\em Proc. NeurIPS}, 2020.

\bibitem{hsu21hubert}
W.~Hsu, B.~Bolte, Y.~H. Tsai, {\em et~al.}, ``{HuBERT: S}elf-supervised speech representation learning by masked prediction of hidden units,'' {\em IEEE Trans. Audio, Speech, Lang. Process.}, 2021.

\bibitem{chen21wavlm}
S.~Chen, C.~Wang, Z.~Chen, {\em et~al.}, ``{WavLM: L}arge-scale self-supervised pre-training for full stack speech processing,'' {\em IEEE J. Sel. Topics Signal Process.}, 2021.

\bibitem{babu22xlsr}
A.~Babu, C.~Wang, A.~Tjandra, {\em et~al.}, ``{XLS-R: S}elf-supervised cross-lingual speech representation learning at scale,'' in {\em Proc. Interspeech}, 2022.

\bibitem{yang21superb}
S.~Yang, P.~Chi, Y.~Chuang, {\em et~al.}, ``{SUPERB: S}peech processing universal performance benchmark,'' in {\em Proc. Interspeech}, 2021.

\bibitem{tsai22superbsg}
H.~Tsai, H.~Chang, W.~Huang, {\em et~al.}, ``{SUPERB-SG: E}nhanced speech processing universal performance benchmark for semantic and generative capabilities,'' in {\em Proc. ACL}, 2022.

\bibitem{shi21mlsuperb}
J.~Shi, D.~Berrebbi, W.~Chen, {\em et~al.}, ``{ML-SUPERB: M}ultilingual speech universal performance benchmark,'' in {\em Proc. Interspeech}, 2023.

\bibitem{huang24dynamic}
C.~Huang, K.~Lu, S.~Wang, {\em et~al.}, ``{Dynamic-SUPERB: T}owards a dynamic, collaborative, and comprehensive instruction-tuning benchmark for speech,'' in {\em Proc. ICASSP}, 2024.

\bibitem{PasadCL21}
A.~Pasad, J.~Chou, and K.~Livescu, ``Layer-wise analysis of a self-supervised speech representation model,'' in {\em Proc. ASRU}, 2021.

\bibitem{PasadSL23}
A.~Pasad, B.~Shi, and K.~Livescu, ``Comparative layer-wise analysis of self-supervised speech models,'' in {\em Proc. ICASSP}, 2023.

\bibitem{choi24mi}
K.~Choi, J.~Jung, and S.~Watanabe, ``Understanding probe behaviors through variational bounds of mutual information,'' in {\em Proc. ICASSP}, 2024.

\bibitem{choi2022opening}
K.~Choi and E.~Yeo, ``Opening the black box of wav2vec feature encoder,'' {\em arXiv preprint arXiv:2210.15386}, 2022.

\bibitem{choi24b_interspeech}
K.~Choi, A.~Pasad, T.~Nakamura, {\em et~al.}, ``Self-supervised speech representations are more phonetic than semantic,'' in {\em Proc. Interspeech}, 2024.

\bibitem{chang2024dsuchallenge}
X.~Chang, J.~Shi, J.~Tian, Y.~Wu, Y.~Tang, Y.~Wu, S.~Watanabe, Y.~Adi, X.~Chen, and Q.~Jin, ``The {Interspeech} 2024 challenge on speech processing using discrete units,'' in {\em Proc. Interspeech}, pp.~2559--2563, 2024.

\bibitem{chang2024exploring}
X.~Chang, B.~Yan, K.~Choi, {\em et~al.}, ``Exploring speech recognition, translation, and understanding with discrete speech units: A comparative study,'' in {\em Proc. ICASSP}, 2024.

\bibitem{borsos2023audiolm}
Z.~Borsos, R.~Marinier, D.~Vincent, {\em et~al.}, ``{AudioLM: A} language modeling approach to audio generation,'' {\em IEEE Trans. Audio, Speech, Lang. Process.}, 2023.

\bibitem{LeeCWGPMPAHTPH22}
A.~Lee, P.~Chen, C.~Wang, {\em et~al.}, ``Direct speech-to-speech translation with discrete units,'' in {\em Proc. ACL}, 2022.

\bibitem{InagumaPKCWC00023}
H.~Inaguma, S.~Popuri, I.~Kulikov, {\em et~al.}, ``{UnitY: T}wo-pass direct speech-to-speech translation with discrete units,'' in {\em Proc. ACL}, 2023.

\bibitem{rubenstein2023audiopalm}
P.~K. Rubenstein, C.~Asawaroengchai, D.~D. Nguyen, {\em et~al.}, ``{AudioPaLM: A} large language model that can speak and listen,'' {\em arXiv preprint arXiv:2306.12925}, 2023.

\bibitem{maiti2024voxtlm}
S.~Maiti, Y.~Peng, S.~Choi, {\em et~al.}, ``{VoxtLM: U}nified decoder-only models for consolidating speech recognition, synthesis and speech, text continuation tasks,'' in {\em Proc. ICASSP}, 2024.

\bibitem{lakhotia2021generative}
K.~Lakhotia, E.~Kharitonov, W.-N. Hsu, {\em et~al.}, ``On generative spoken language modeling from raw audio,'' {\em Trans. ACL}, 2021.

\bibitem{hassid2024textually}
M.~Hassid, T.~Remez, T.~A. Nguyen, {\em et~al.}, ``Textually pretrained speech language models,'' in {\em Proc. NeurIPS}, 2024.

\bibitem{van2022comparison}
B.~Van~Niekerk, M.-A. Carbonneau, J.~Za{\"\i}di, {\em et~al.}, ``A comparison of discrete and soft speech units for improved voice conversion,'' in {\em Proc. ICASSP}, 2022.

\bibitem{liang2014improved}
Y.~Liang, M.-F. Balcan, V.~Kanchanapally, and D.~Woodruff, ``Improved distributed principal component analysis,'' in {\em Proc. NeurIPS}, 2014.

\bibitem{sanabria2023analyzing}
R.~Sanabria, H.~Tang, and S.~Goldwater, ``Analyzing acoustic word embeddings from pre-trained self-supervised speech models,'' in {\em Proc. ICASSP}, 2023.

\bibitem{tom2022wav2vec}
T.~tom Dieck, P.~A. P{\'e}rez-Toro, T.~Arias, {\em et~al.}, ``Wav2vec behind the scenes: How end2end models learn phonetics.,'' in {\em Proc. Interspeech}, 2022.

\bibitem{mohamed2024orthogonality}
M.~Mohamed, O.~D. Liu, H.~Tang, and S.~Goldwater, ``Orthogonality and isotropy of speaker and phonetic information in self-supervised speech representations,'' in {\em Proc. Interspeech}, 2024.

\bibitem{YamagiwaOS23}
H.~Yamagiwa, M.~Oyama, and H.~Shimodaira, ``Discovering universal geometry in embeddings with {ICA},'' in {\em Proc. EMNLP}, 2023.

\bibitem{chung2021w2vbert}
Y.-A. Chung, Y.~Zhang, W.~Han, {\em et~al.}, ``{w2v-BERT: C}ombining contrastive learning and masked language modeling for self-supervised speech pre-training,'' in {\em Proc. ASRU}, pp.~244--250, 2021.

\bibitem{su2021whitening}
J.~Su, J.~Cao, W.~Liu, and Y.~Ou, ``Whitening sentence representations for better semantics and faster retrieval,'' {\em arXiv preprint arXiv:2103.15316}, 2021.

\bibitem{huang2021whiteningbert}
J.~Huang, D.~Tang, W.~Zhong, {\em et~al.}, ``{W}hitening{BERT}: An easy unsupervised sentence embedding approach,'' in {\em Proc. EMNLP}, 2021.

\bibitem{sawada2019apsipatrans}
H.~Sawada, N.~Ono, H.~Kameoka, {\em et~al.}, ``A review of blind source separation methods{: Two} two converging routes to {ILRMA} originating from {ICA} and {NMF},'' {\em APSIPA Trans. Signal Inf. Process.}, vol.~8, no.~e12, pp.~1--14, 2019.

\bibitem{cardoso1997spl}
J.-F. Cardoso, ``Infomax and maximum likelihood for blind source separation,'' {\em IEEE Signal Process. Lett.}, vol.~4, no.~4, pp.~112--114, 1997.

\bibitem{Ono2010LVAICA}
N.~Ono and S.~Miyabe, ``Auxiliary-function-based independent component analysis for super-gaussian sources,'' in {\em Proc. LVA/ICA}, pp.~165--172, 2010.

\bibitem{panayotov2015icassp}
V.~Panayotov, G.~Chen, D.~Povey, and S.~Khudanpur, ``{LibriSpeech: A}n {ASR} corpus based on public domain audio books,'' in {\em Proc. ICASSP}, pp.~5206--5210, 2015.

\bibitem{li2024dsu}
Z.~Li, Y.~Yang, X.~Li, J.~Kang, X.-L. Zhang, and J.~Li, ``The {ChinaTelecom I}nterspeech2024 discrete speech unit asr challenge system description,'' in {\em Submission for {Interspeech} 2024 Challenge on Speech Processing Using Discrete Units}, 2024.

\bibitem{chang23b_interspeech}
X.~Chang, B.~Yan, Y.~Fujita, T.~Maekaku, and S.~Watanabe, ``Exploration of efficient end-to-end {ASR} using discretized input from self-supervised learning,'' in {\em Proc. Interspeech}, pp.~1399--1403, 2023.

\bibitem{kim2023slt}
K.~Kim, F.~Wu, Y.~Peng, J.~Pan, P.~Sridhar, K.~J. Han, and S.~Watanabe, ``{E-Branchformer: B}ranchformer with enhanced merging for speech recognition,'' pp.~84--91, 2023.

\bibitem{GodeyCS24}
N.~Godey, {\'{E}}.~V. de~la Clergerie, and B.~Sagot, ``Anisotropy is inherent to self-attention in transformers,'' 2024.

\bibitem{wu2024codec}
H.~Wu, X.~Chen, Y.-C. Lin, {\em et~al.}, ``Codec-superb@ slt 2024: A lightweight benchmark for neural audio codec models,'' in {\em Proc. SLT}, 2024.

\bibitem{lee2022naacl}
A.~Lee, H.~Gong, P.-A. Duquenne, H.~Schwenk, P.-J. Chen, C.~Wang, S.~Popuri, Y.~Adi, J.~Pino, J.~Gu, and W.-N. Hsu, ``Textless speech-to-speech translation on real data,'' in {\em Proc. NAACL HLT}, pp.~860--872, 2022.

\bibitem{mu2018all}
J.~Mu and P.~Viswanath, ``All-but-the-top: Simple and effective postprocessing for word representations,'' in {\em Proc. ICLR}, 2018.

\bibitem{raunak2019effective}
V.~Raunak, V.~Gupta, and F.~Metze, ``Effective dimensionality reduction for word embeddings,'' in {\em Proc. RepL4NLP@ACL}, 2019.

\bibitem{garofolo1993darpa}
J.~Garofolo, L.~Lamel, W.~Fisher, {\em et~al.}, ``{DARPA TIMIT: A}coustic-phonetic continuous speech corpus {CD-ROM, NIST} speech disc 1-1.1,'' 1993.

\bibitem{abdullah23_interspeech}
B.~M. Abdullah, M.~M. Shaik, B.~Möbius, and D.~Klakow, ``An information-theoretic analysis of self-supervised discrete representations of speech,'' in {\em Proc. Interspeech}, 2023.

\bibitem{sicherman2023analysing}
A.~Sicherman and Y.~Adi, ``Analysing discrete self supervised speech representation for spoken language modeling,'' in {\em Proc. ICASSP}, 2023.

\end{thebibliography}

\end{document}